\documentclass[runningheads]{llncs}
\usepackage[T1]{fontenc}
\usepackage{adjustbox}
\usepackage{multirow}
\usepackage{mathtools}
\usepackage{graphicx}
\usepackage[dvipsnames]{xcolor}
\usepackage[title]{appendix}%
\usepackage[pagebackref=true,breaklinks=true,colorlinks,bookmarks=false]{hyperref}
\begin{document}
\title{
\begin{minipage}{0.1\textwidth}
\includegraphics[width=\linewidth]{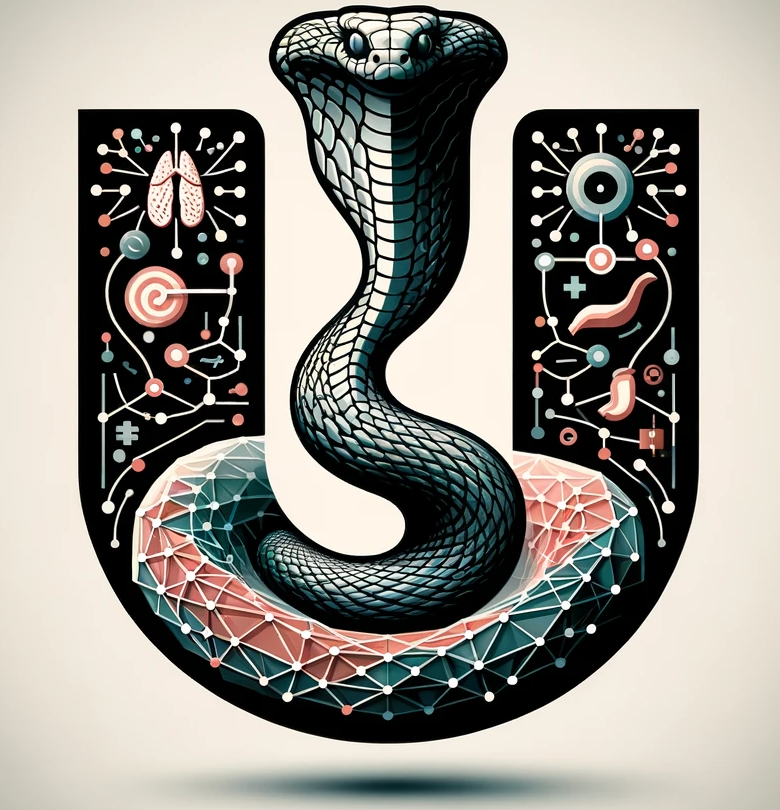}
\end{minipage}
U-Mamba: Enhancing Long-range\\ Dependency for
Biomedical Image Segmentation}

\titlerunning{U-Mamba}

\author{Jun Ma \inst{1,2,3} \and
Feifei Li \inst{1} \and 
Bo Wang \inst{1,2,3,4,5}
} 
\authorrunning{J. Ma, F. Li, and B. Wang}

\institute{Peter Munk Cardiac Centre, University Health Network, Toronto, Canada \and
Department of Laboratory Medicine and Pathobiology, University of Toronto, Toronto, Canada \and
Vector Institute for Artificial Intelligence, Toronto, Canada \and
Department of Computer Science, University of Toronto, Toronto, Canada \and
AI Hub, University Health Network, Toronto, Canada \\
\email{bowang@vectorinstitute.ai}}
\maketitle              % typeset the header of the contribution
\begin{abstract}
Convolutional Neural Networks (CNNs) and Transformers have been the most popular architectures for biomedical image segmentation, but both of them have limited ability to handle long-range dependencies because of inherent locality or computational complexity. 
To address this challenge, we introduce U-Mamba, a general-purpose network for biomedical image segmentation.
Inspired by the State Space Sequence Models (SSMs), a new family of deep sequence models known for their strong capability in handling long sequences, 
we design a hybrid CNN-SSM block that integrates the local feature extraction power of convolutional layers with the abilities of SSMs for capturing the long-range dependency. 
Moreover, U-Mamba enjoys a self-configuring mechanism, allowing it to automatically adapt to various datasets without manual intervention.
We conduct extensive experiments on four diverse tasks, including the 3D abdominal organ segmentation in CT and MR images, instrument segmentation in endoscopy images, and cell segmentation in microscopy images. The results reveal that U-Mamba outperforms state-of-the-art CNN-based and Transformer-based segmentation networks across all tasks. 
This opens new avenues for efficient long-range dependency modeling in biomedical image analysis. 
The code, models, and data are publicly available at \url{https://wanglab.ai/u-mamba.html}.

\keywords{Image Segmentation \and Backbone Network \and State Space Models \and Transformer.}
\end{abstract}

\section{Introduction}
% P1. Segmentation and methodology development background
Segmentation serves as a foundational process in biomedical image analysis that divides images into meaningful anatomical structures or areas of interest for doctors and biologists. This task is crucial in various biomedical applications~\cite{seg-reviewPAMI,BioSegFM-NM23}, such as diagnosing diseases, quantifying cancer microenvironments, planning treatment strategies, and tracking disease progression. Manual segmentation is a straightforward way for this task but it is extremely time-consuming and requires extensive domain knowledge, which is impractical in practice. Thus, there is a great demand for automatic segmentation methods. During the past ten years, segmentation 
methodology has undergone a paradigm shift from Mathematical model- or hand-crafted feature-based approaches to deep learning-based paradigm, which significantly improved the segmentation accuracy and efficiency for a broad range of biomedical tasks, such as tumor segmentation~\cite{LiTS,KiTS} and organ segmentation~\cite{FLARE22} in 3D Computed Tomograph (CT) scans, as well as cell segmentation in microscopy images~\cite{cellpose,NeurIPS22CellSeg}.

Two popular network architectures, CNNs~\cite{CNN95} and Transformers~\cite{attention-Nips17}, have gained prominence in medical image segmentation. CNNs, such as U-Net~\cite{U-Net,nnUNet} and DeepLab~\cite{deeplabV3plus}, are efficient in extracting hierarchical image features and more parameter-efficient than na\"ive fully connected networks. Their shared weights architecture makes them excel at capturing translational invariances and recognizing local features. Transformers were originally designed for natural language processing, but have been successfully adapted for image processing, such as Vision Transformer (ViT)~\cite{ViT2020} for image recognition and SwinTransformer~\cite{Swin} for a general backbone in various vision tasks. In contrast to CNNs, Transformers do not inherently process spatial image hierarchies but treat the image as a sequence of patches. which has better capabilities to capture global information. Due to this complementary feature, many studies have explored incorporating Transformers into CNNs via hybrid network architectures, such as TransUNet~\cite{TransUNet,TransUNet3D}, UNETR~\cite{UNETR}, nnFormer~\cite{nnFormer}, and SwinUNETR~\cite{swinunetr}.

While Transformers have improved the ability of long-range dependencies, they are generally very computationally expensive. This is because the self-attention mechanism scales quadratically with the input size, making them resource-intensive, especially for biomedical images that usually have high resolutions. Thus, how to efficiently enhance the long-range dependency in CNNs remains an open question. 
Recently, state space sequence models (SSMs)~\cite{GuThesis,SSM-NeurIPS21}, in particular structured state space sequence models (S4)~\cite{S4-ICLR22}, have emerged as an efficient and effective building block (or layer) for constructing deep networks, which obtained cutting-edge performance in continuous long-sequence data analysis~\cite{SSM4Audio,S4-ICLR22}. 
Mamba~\cite{mamba} further improved S4 with a selective mechanism, allowing the model to select relevant information in an input-dependent manner. By combining with hardware-aware implementation, Mamba surpassed Transformers on dense modalities such as language and genomics.
On the other hand, state space models also have shown promising results on vision tasks, such as images~\cite{S4ND-neurips22} and videos~\cite{SSM4Video-ECCV22} classification. 
Since image patches and image features can be cast as sequences~\cite{ViT2020,Swin}, these appealing features of SSMs motivated us to explore the potential of using Mamba blocks to enhance the long-range modeling ability in CNNs.

% contributions 
In this paper, we propose U-Mamba, a general-purpose network, for both 3D and 2D biomedical image segmentation. Based on the innovative hybrid CNN-SSM architecture, U-Mamba can capture both localized fine-grained features and long-range dependencies in images. This network stands out from prevalent Transformer-based architectures by offering linear scaling in feature size, as opposed to the quadratic complexity often associated with Transformers. Additionally, U-Mamba's self-configuring capability enables seamless adaptation to various datasets, enhancing its scalability and flexibility across a spectrum of biomedical segmentation tasks.
Quantitative and qualitative results on four distinct datasets demonstrate that U-Mamba achieves superior performance, surpassing Transformer-based networks by a large margin. This paves the way for future advancements in network designs that efficiently and effectively model long-range dependencies in biomedical imaging.

\section{Method}
\begin{figure}[htbp]
\centering
\includegraphics[scale=0.11]{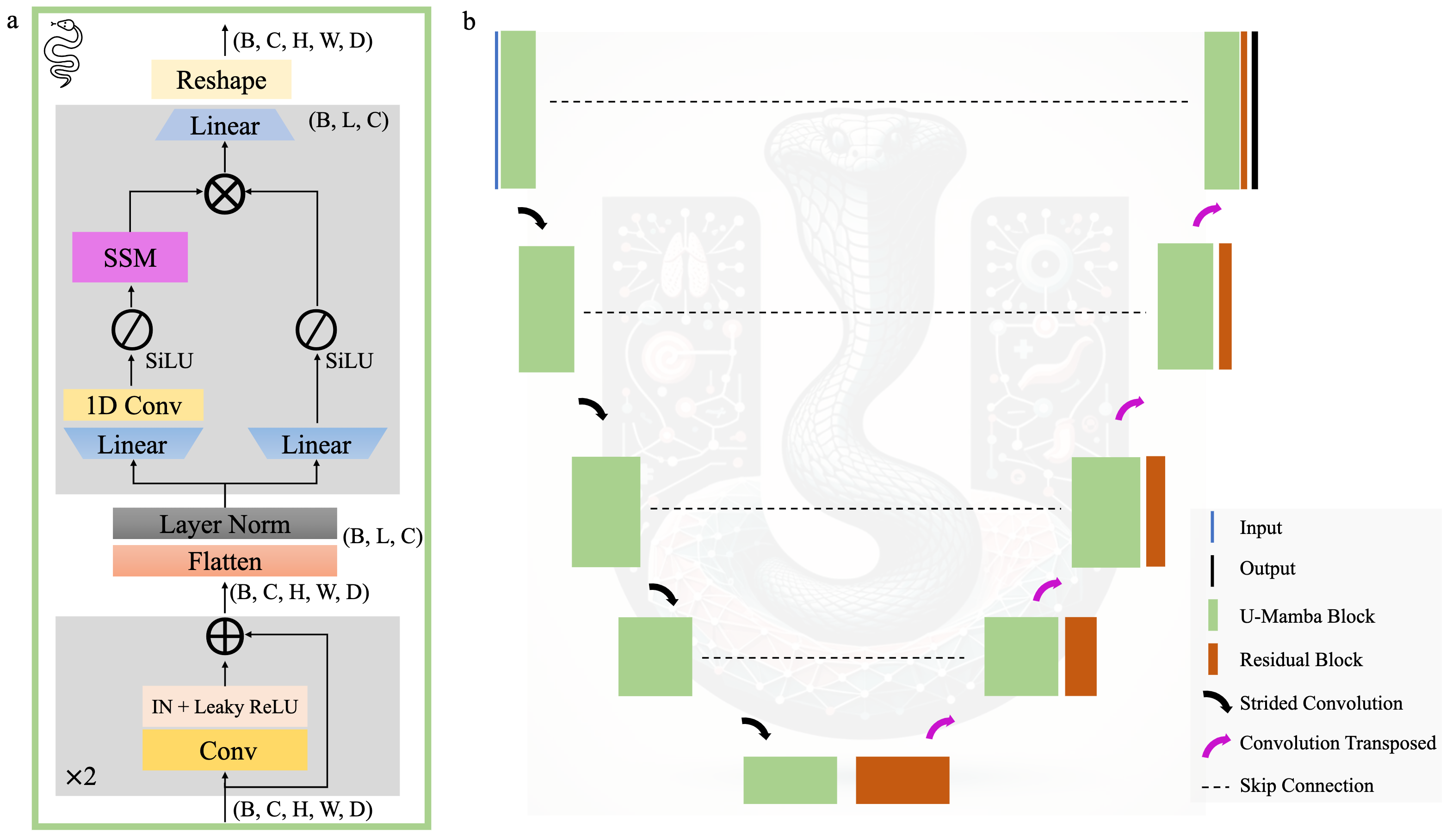}
\caption{Overview of the U-Mamba (Enc) architecture. 
\textbf{a,} U-Mamba building block contains two successive Residual blocks followed by the SSM-based Mamba block for enhanced long-range dependency modeling. 
\textbf{b,} U-Mamba employs the encoder-decoder framework with U-Mamba blocks in the endocer, Residual blocks in the decoder, together with skip connections. 
Note: This illustration serves as a conceptual representation.  U-Mamba inherits the self-configuring feature from nnU-Net and the number of network blocks is automatically determined across datasets. The detailed network configurations are presented in Table~\ref{tab:config}. 
}
\label{fig:network}
\end{figure}

U-Mamba follows the encoder-decoder network structure that captures both local features and long-range contexts in an efficient way. Fig.~\ref{fig:network} shows an overview of the U-Mamba block and the complete network structure. 
Next, we first introduce the Mamba block followed by illustrating the details of U-Mamba. 

\subsection{Mamba: Selective Structured State Space Sequence Models with a Scan (S6)}
State Space Sequence Models (SSMs)~\cite{GuThesis} are a type of systems that map a 1-dimensional function or sequence $u(t) \mapsto y(t) \in \mathrm{R}$ and can be represented as the following linear Ordinary Differential Equation (ODE): 
\begin{equation}
\begin{aligned}
    x'(t) &= \mathbf{A}x(t) + \mathbf{B}u(t), \\
    y(t) &= \mathbf{C}x(t),
\end{aligned}
\end{equation}
where state matrix $A\in \mathrm{R}^{N\times N}$ and $B,C\in \mathrm{R}^N$ are its parameters and $x(t) \in \mathrm{R}^N$ denotes the implicit latent state. 
SSMs offer several desired properties, such as linear computational complexity per time step and parallelized computation for efficient training. However, na\"ive SSMs typically demand more memory than equivalent CNNs and often encounter vanishing gradients during training, hindering their broader application in general sequence modeling.

Structured State Space Sequence Models (S4)~\cite{S4-ICLR22} significantly improve the na\"ive SSMs by imposing structured forms on the state matrix $\mathbf{A}$ and introducing an effective algorithm. Specifically, the state matrix is crafted and initialized with High-Order Polynomial Projection Operator (HIPPO)~\cite{gu2020hippo}, allowing to building deep sequence models with rich capability and efficient long-range reasoning ability. S4, as a new type of network architecture, has surpassed Transformers~\cite{attention-Nips17} on the challenging Long Range Arena Benchmark~\cite{longRangeArena} by a significant margin.

Recently, Mamba~\cite{mamba} further advances SSMs in discrete data modeling (e.g., text and genome) through two key improvements. First, Mamba employs an input-dependent selection mechanism, distinct from the traditional time- and input-invariant SSMs, which allows for efficient information filtering from inputs. This is implemented by parameterizing the SSM parameters based on the input data.
Second, a hardware-aware algorithm, scaling linearly in sequence length, is developed to compute the model recurrently with a scan, making Mamba faster than previous methods on modern hardware. In addition, The Mamba architecture, merging SSM blocks with linear layers, is notably simpler and has demonstrated state-of-the-art performance in various long-sequence domains including language and genomics, showcasing significant computational efficiency during both training and inference.

\subsection{U-Mamba: Marry Mamba with U-Net}
Mamba has demonstrated impressive results on various discrete data, yet its potential for modeling image data, particularly in biomedical imaging, remains underexplored.  Given that images are essentially discrete data sampled from continuous signals and can be treated as long sequences when flattened, we propose leveraging Mamba's linear scaling advantage to enhance CNNs' long-range dependency modeling. 
Transformers have been successfully applied to handle images, such as ViT~\cite{ViT2020} and SwinTransformer~\cite{Swin}, but they often suffer from high computational burden for large images because self-attention~\cite{attention-Nips17} has quadratic complexity. These factors motivate us to leverage the linear scaling property in Mamba to enhance the long-range dependency of CNNs.

U-Net~\cite{U-Net} and its variants~\cite{nnUNet} have been the widely used network architecture in medical image segmentation. They usually have symmetric encoder-decoder architecture to extract multi-scale image features with convolutional operations. However, this architectural design has limited ability to model long-range dependencies in images because convolutional kernels are inherently local. Each convolutional layer captures features from a limited receptive field. While skip connections in U-Net help in combining low-level details with high-level features, they primarily reinforce local feature integration rather than expanding the network's ability to model long-range dependencies.

U-Mamba is designed to acquire the best of U-Net and Mamba for global context understanding in medical image segmentation. As shown in Fig.~\ref{fig:network}a, each building block contains two successive Residual blocks~\cite{resnet} followed by a Mamba block~\cite{mamba}. The Residual block contains the plain convolutional layer followed by Instance Normalization (IN)~\cite{instance-norm} and Leaky ReLU~\cite{leakyRelu}. Image features with a shape of $(B, C, H, W, D)$ are then flattened and transposed to $(B, L, C)$ where $L=H\times W \times D$. After passing the Layer Normalization~\cite{layerNorm}, the features enter the Mamba block that contains two parallel branches. In the first branch, the features are expanded to $(B, 2L, C)$ by a linear layer followed by a 1D convolutional layer, SiLU activation function~\cite{silu}, and together with the SSM layer. In the second branch, the features are also expanded to $(B, 2L, C)$ with a linear layer followed by the SiLU activation function. After that, the features from the two branches are merged with the Hadamard product. Finally, the features are projected back to the original shape $(B, L, C)$ and then reshaped and transposed to $(B, C, H, W, D)$.

Fig.~\ref{fig:network}b shows the complete U-Mamba network architecture where the encoder is built with the above blocks to capture both local features and long-range dependencies.
The decoder, composed of Residual blocks and transposed convolutions, focuses on detailed local information and resolution recovery. 
Moreover, we inherit the skip connection in U-Net to connect the hierarchical features from the encoder to the decoder. The final decoder feature is passed to a $1\times 1 \times 1$ convolutional layer together with a Softmax layer to produce the final segmentation probability map.
In addition, we also implement a U-Mamba variant where only the bottleneck uses the U-Mamba block and the others are plain Residual blocks. We use ``U-Mamba\_Bot'' and ``U-Mamba\_Enc'' to differentiate the two network variants, which employ the U-Mamba block in the bottleneck and all encoder blocks, respectively.

\section{Experiments and results}
\subsection{Datasets}
We used four medical image datasets to evaluate the performance and scalability of U-Mamba across different image sizes, segmentation targets, and modalities. Table~\ref{tab:data} presents an overview of the datasets. All the datasets are publicly available and allowed for research purposes.

\begin{table}[htbp]
\caption{Dataset information.}\label{tab:data}
\centering
\begin{tabular}{lcccc}
\hline
Dataset    & Dimension &  \#Training Image & \#Testing Image & \#Targets \\ \hline
Abdomen CT & 3D   & 50  (4794 slices)            & 50 (10894 slices)            &   13          \\
Abdomen MRI & 3D   & 60  (5615 slices)            & 50  (3357 slices)           &   13          \\
Endoscopy Images & 2D   & 1800           & 1200           &  7           \\
Microscopy Images & 2D   & 1000           & 101           &  2           \\ \hline
\end{tabular}
\end{table}

\textbf{Abdomen CT:} This dataset was from the MICCAI 2022 FLARE Challenge~\cite{FLARE22} that focused on the segmentation of 13 abdominal organs, including the liver, spleen, pancreas, right kidney, left kidney, stomach, gallbladder, esophagus, aorta, inferior vena cava, right adrenal gland, left adrenal gland, and duodenum. The training set contained 50 CT scans that were from the MSD Pancreas dataset~\cite{simpson2019MSD} and the annotations were from AbdomenCT-1K. Another 50 cases from different medical centers~\cite{TCIA} were used for evaluation and the annotations were provided by the challenge organizers.

\textbf{Abdomen MRI:} This dataset was from the MICCAI 2022 AMOS Challenge~\cite{AMOS22} that also focused on abdominal organ segmentation. The original dataset contains 40 and 20 MRI scans for training and validation, respectively. Since 20 cases are not enough to draw statistically meaningful results, in our experiments, we used the original 60 labeled MRI scans for model training and annotated another 50 MRI scans as the testing set. In order to enable modality-wise comparison of abdominal organs, we also focused on the same 13 organs as the Abdomen CT dataset.
The annotations were generated by radiologists with the assistance of MedSAM~\cite{MedSAM} and ITK-SNAP~\cite{ITKSNAP}. We have released this new annotated dataset to the community to promote the development of abdominal organ segmentation in MRI.

\textbf{Endoscopy images:} This dataset was from the MICCAI 2017 EndoVis Challenge~\cite{Endo17} that focused on segmenting seven instruments from endoscopy images, including the large needle driver, prograsp forceps, monopolar
curved scissors, cadiere forceps, bipolar forceps, vessel sealer, and additionally a drop-in ultrasound probe. We followed the official dataset splitting where the training set contained 1800 and 1200 image frames, respectively. The training images were from eight videos while the testing set had unseen images from another two new videos.

\textbf{Microscopy images:} This dataset was from the NeurIPS 2022 Cell Segmentation Challenge~\cite{NeurIPS22CellSeg} that focused on cell segmentation in various microscopy images. We used 1000 and 101 images for training and evaluation, respectively. Different from the above three tasks, this is an instance segmentation task where algorithms are expected to assign a unique label for each cell instance. In our experiments, we converted the instance segmentation into a semantic segmentation task by predicting the cell boundaries and interior regions because they can be easily transformed to instance masks via the ``skimage.measure.label'' function.
It should be noted that our purpose was to benchmark the performance of network architectures, rather than pursue state-of-the-art performance on this task. U-Mamba can also serve as a backbone network in state-of-the-art instance segmentation framework~\cite{NeurIPS22CellSeg}. We leave this extension as future work.

\subsection{Implementation and training protocols}
We implemented U-Mamba within the popular nnU-Net~\cite{nnUNet} framework driven by two desired features. 
First, the modular design of nnU-Net aligns perfectly with our focus on introducing a new network architecture. This design allows us to concentrate on the network's implementation while controlling other variables, such as image preprocessing and data augmentation. Such a setup enables a fair comparison of U-Mamba against various methods under uniform conditions, with the network architecture being the only differing factor.
Second, nnU-Net's notable self-configuring framework, capable of automatically configuing hyper-parameters for diverse segmentation datasets, is another compelling reason for our choice. 
We also preserve this feature, allowing U-Mamba to be easily adapted to a broad range of segmentation tasks. During training, the patch size, batch size, and network configurations (e.g., the number of resolution states and the number of downsampling operations along different axes) remained consistent with nnU-Net (Table~\ref{tab:config}). U-Mamba was also optimized with stochastic gradient descent and the loss function was the unweighted sum of Dice loss and cross-entropy because the compound loss has proven to be robust across different tasks~\cite{LossOdyssey}.
During inference, the testing time augmentation (TTA) was disabled in all the experiments for a more streamlined and efficient evaluation process because TTA would increase the computational burden by $4\times$ and $8\times$ times for 2D and 3D datasets, respectively.

\begin{table}[htbp]
\caption{U-Mamba configurations for each dataset.}\label{tab:config}
\centering
\begin{tabular}{lcccc}
\hline
Configurations & Patch size     & Batch size & \# Stages & \# Pooling per axis \\ \hline
Abdomen CT     & (40, 224, 192) & 2          & 6           & (3, 3, 5)             \\
3D Abdomen MR  & (48, 160, 224) & 2          & 6           & (3, 5, 5)             \\
2D Abdomen MR  & (320, 320)     & 30         & 7           & (6, 6)                \\
Endoscopy      & (384, 640)     & 13         & 7           & (6, 6)                \\
Microscopy     & (512, 512)     & 12         & 8           & (7, 7)                \\ \hline
\end{tabular}
\end{table}

\subsection{Benchmarking}
We compared U-Mamba with two CNN-based segmentation networks (nnU-Net~\cite{nnUNet} and SegResNet~\cite{SegResNet}) and two Transformer-based networks (UNETR~\cite{UNETR} and SwinUNETR~\cite{swinunetr}), which are widely used in medical image segmentation competitions. 
% Notably, nnU-Net and SegResNet have been proposed for several years, but they were still the core network architecture of many winning solutions in MICCAI 2022-23 segmentation challenges. 
For a fair comparison, we also implemented SegResNet, UNETR, and SwinUNETR into the nnU-Net framework and used their suggested optimizers (e.g., Adam~\cite{ADAM15} and AdamW~\cite{adamW}) for model training. We used the default image preprocessing in nnU-Net~\cite{nnUNet}.
All the networks were trained from scratch for 1000 epochs on one NVIDIA A100 GPU in the same batch size (Table~\ref{tab:config}).

Following the recommendations in Metrics Reloaded~\cite{metrics-reloaded}, we used Dice Similarity Coefficient (DSC) and Normalized Surface Distance (NSD) for three semantic segmentation tasks, including organ segmentation in CT and MRI scans, and instrument segmentation in Endoscopy images. F1 score was used to evaluate the cell segmentation quality because it is an instance segmentation task.

\begin{table}[htb]
\caption{Results summary of 3D organ segmentation on abdomen CT and MRI datasets. U-Mamba\_Bot: only use the U-Mmaba block in the bottleneck.  U-Mamba\_Enc: all encoder blocks are U-Mmaba blocks.}\label{tab:results-3d}
\centering
\begin{adjustbox}{width=0.99\textwidth}
\begin{tabular}{lcc|cc}
\hline
\multirow{2}{*}{Methods} & \multicolumn{2}{c|}{Organs in Abdomen CT}                 & \multicolumn{2}{c}{Organs in Abdomen MRI}                 \\ \cline{2-5} 
                         & DSC                    & NSD                    & DSC                    & NSD                    \\ \hline
nnU-Net                  & 0.8615$\pm$0.0790          & 0.8972$\pm$0.0824          & 0.8309$\pm$0.0769          & 0.8996$\pm$0.0729          \\
SegResNet                & 0.7927$\pm$0.1162          & 0.8257$\pm$0.1194          & 0.8146$\pm$0.0959          & 0.8841$\pm$0.0917          \\
UNETR                    & 0.6824$\pm$0.1506          & 0.7004$\pm$0.1577          & 0.6867$\pm$0.1488          & 0.7440$\pm$0.1627          \\
SwinUNETR                & 0.7594$\pm$0.1095          & 0.7663$\pm$0.1190          & 0.7565$\pm$0.1394          & 0.8218$\pm$0.1409          \\ \hline
U-Mamba\_Bot             & \textbf{0.8683$\pm$0.0808} & \textbf{0.9049$\pm$0.0821} & \textbf{0.8453$\pm$0.0673} & \textbf{0.9121$\pm$0.0634} \\
U-Mamba\_Enc             & \textbf{0.8638$\pm$0.0908} & \textbf{0.8980$\pm$0.0921} & \textbf{0.8501$\pm$0.0732} & \textbf{0.9171$\pm$0.0689} \\ \hline
\end{tabular}
\end{adjustbox}
\end{table}

\begin{figure}[!htp]
\centering
\includegraphics[scale=0.08]{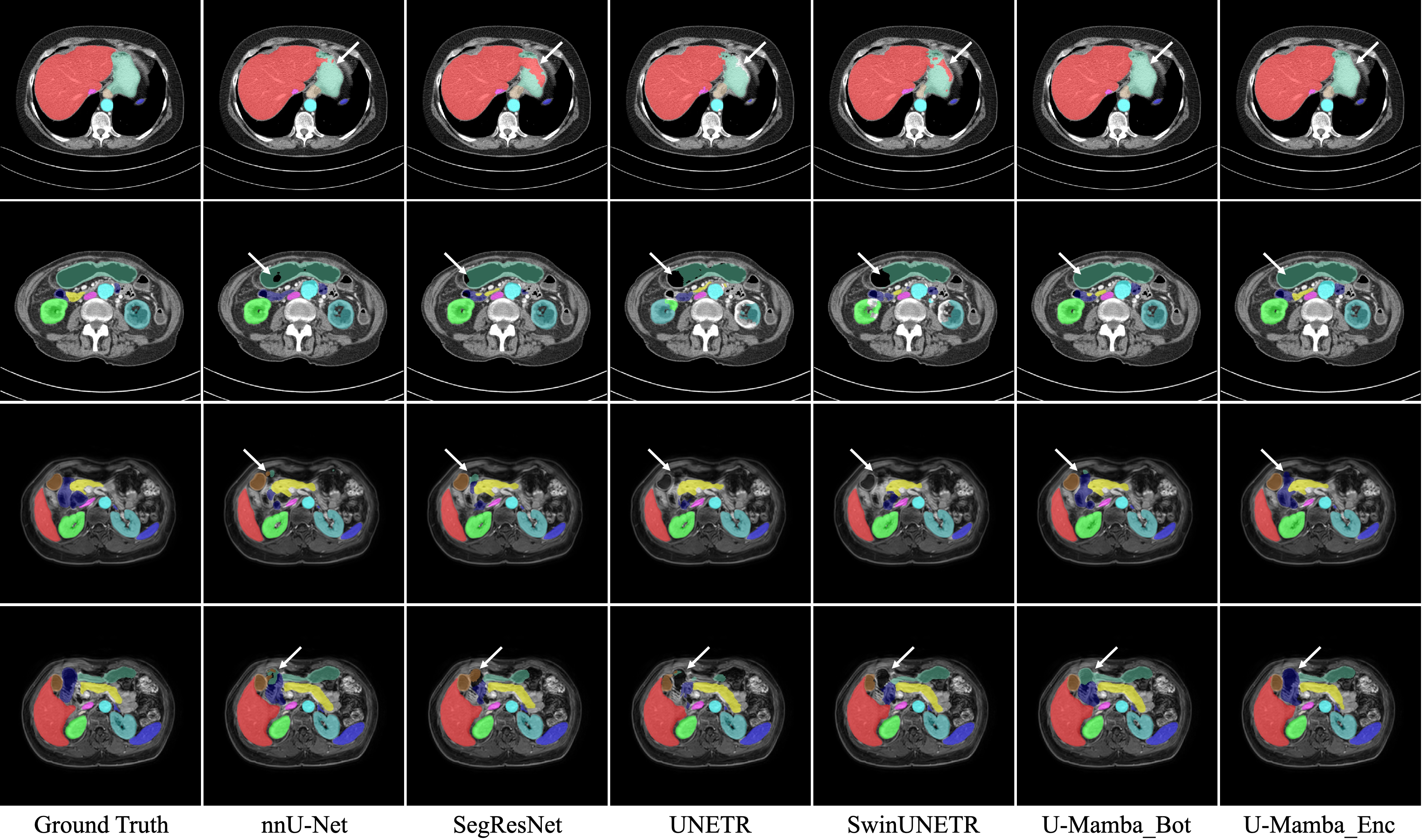}
\caption{Visualized segmentation examples of abdominal organ segmentation in CT (1st and 2nd rows) and MRI scans (3rd and 4th rows). U-Mamba has better capabilities to distinguish complex soft tissues in the abdomen. 
}
\label{fig:results-3d}
\end{figure}

\newpage
\subsection{Quantitative and qualitative segmentation results}

Table~\ref{tab:results-3d} shows quantitative 3D segmentation results of abdominal organs in CT and MRI scans.
U-Mamba surpasses both CNN and Transformer-based segmentation networks, achieving average DSC scores of 0.8683 (U-Mamba\_Bot) and 0.8501 (U-Mamba\_Enc) on the abdomen CT and MR datasets, respectively. nnU-Net performs competitively, which is much better than SegResNet, UNETR, and SwinUNETR, but U-Mamba still shows advantages as shown in Fig.~\ref{fig:results-3d}. Based on the visualized segmentation results of the six networks, U-Mamba has fewer outliers in the segmentation results. For instance, U-Mama can generate more accurate liver and stomach segmentation masks while the others have over-segmentation errors for liver masks and missing regions for stomach masks in abdomen CT scans. Similarly, for gallbladder segmentation in MRI scans, U-Mamba successfully delineates its boundary while other methods generate various segmentation errors. These observations are also reflected in the organ-wise DSC and NSD scores (Appendix Table~\ref{tab:organ-3dCT} and Table~\ref{tab:organ-3dMRI}).

\begin{table}[htb]
\caption{Results summary of 2D segmentation tasks: organ segmentation in abdomen MRI scans, instruments segmentation in endoscopy images, and cell segmentation in microscopy images.}\label{tab:results-2d}
\centering
\begin{adjustbox}{width=0.99\textwidth}
\begin{tabular}{lcc|ll|c}
\hline
\multirow{2}{*}{Methods} & \multicolumn{2}{c|}{Organs in Abdomem MRI}              & \multicolumn{2}{c|}{Instruments in Endoscopy}           & Cells in Microscopy        \\ \cline{2-6} 
                         & DSC                        & NSD                        & \multicolumn{1}{c}{DSC}    & \multicolumn{1}{c|}{NSD}   & F1                         \\ \hline
nnU-Net                  & 0.7450$\pm$0.1117          & 0.8153$\pm$0.1145          & 0.6264$\pm$0.3024          & 0.6412$\pm$0.3074          & 0.5383$\pm$0.2657          \\
SegResNet                & 0.7317$\pm$0.1379          & 0.8034$\pm$0.1386          & 0.5820$\pm$0.3268          & 0.5968$\pm$0.3303          & 0.5411$\pm$0.2633          \\
UNETR                    & 0.5747$\pm$0.1672          & 0.6309$\pm$0.1858          & 0.5017$\pm$0.3201          & 0.5168$\pm$0.3235          & 0.4357$\pm$0.2572          \\
SwinUNETR                & 0.7028$\pm$0.1348          & 0.7669$\pm$0.1442          & 0.5528$\pm$0.3089          & 0.5683$\pm$0.3123          & 0.3967$\pm$0.2621          \\ \hline
U-Mamba\_Bot             & \textbf{0.7588$\pm$0.1051} & \textbf{0.8285$\pm$0.1074} & \textbf{0.6540$\pm$0.3008} & \textbf{0.6692$\pm$0.3050} & \textbf{0.5389$\pm$0.2817} \\
U-Mamba\_Enc             & \textbf{0.7625$\pm$0.1082} & \textbf{0.8327$\pm$0.1087} & \textbf{0.6303$\pm$0.3067} & \textbf{0.6451$\pm$0.3104} & \textbf{0.5607$\pm$0.2784} \\ \hline
\end{tabular}
\end{adjustbox}
\end{table}

\begin{figure}[!htp]
\centering
\includegraphics[scale=0.09]{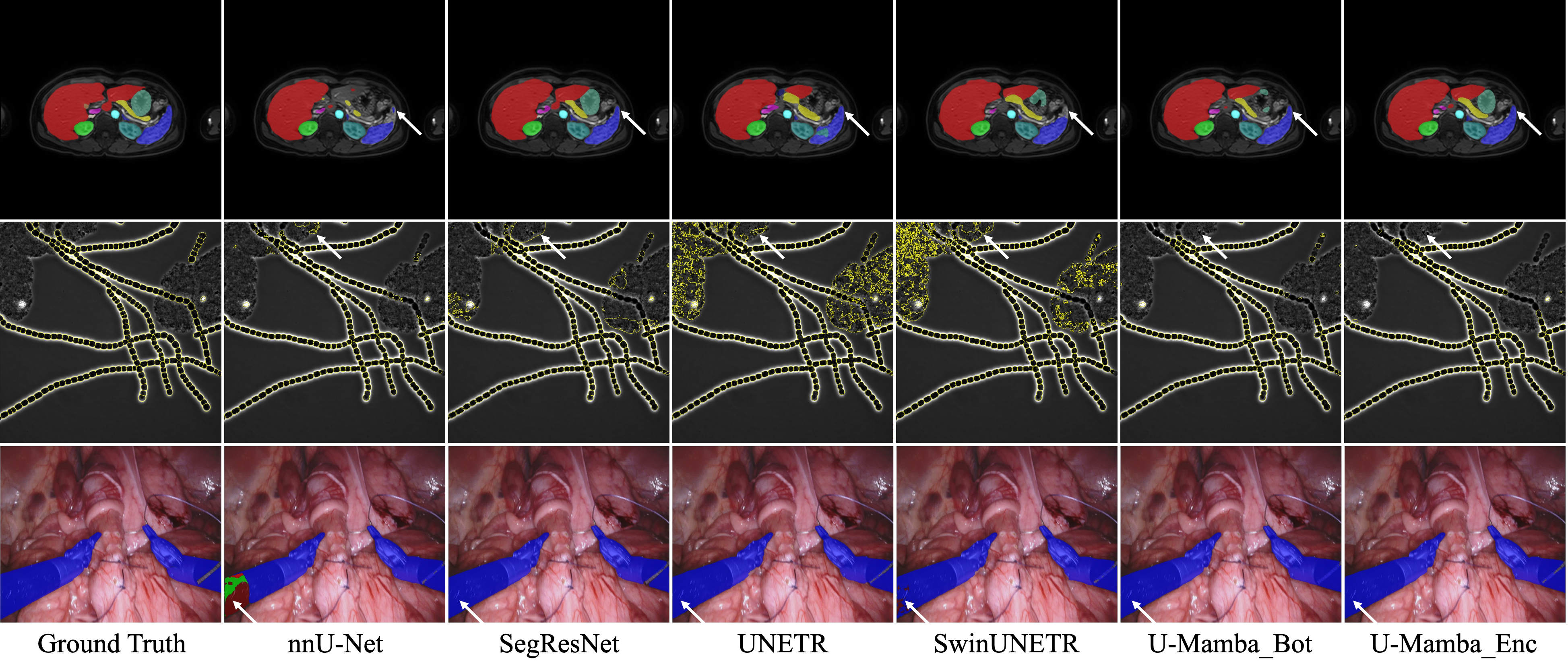}
\caption{Visualized segmentation examples of abdominal organ segmentation MRI scans (1st row), cell segmentation in microscopy images (2nd row), and instruments segmentation in endoscopy images (3rd row). U-Mamba is more robust to heterogeneous appearances and has fewer segmentation outliers. 
}
\label{fig:results-2d}
\end{figure}

Table~\ref{tab:results-2d} presents quantitative 2D segmentation results. 
MRI scans usually have more heterogeneous spacing than CT scans and 2D segmentation networks are also a common choice in practice~\cite{MMs}. Thus, we also compared all the networks under the 2D setting for abdominal organ segmentation where each 3D MRI scan was converted into multiple 2D slices. U-Mamba again demonstrates its advantages over existing methods on the three 2D segmentation tasks, achieving the best average DSC of 0.7625, 0.6504, and F1 score of 0.5607 for organs, instruments, and cell segmentation, respectively. Qualitative results in Fig.~\ref{fig:results-3d} show that nnU-Net, UNETR, and SwinUNETR confuse the spleen because of heterogeneous appearances. UNETR and SwinUNETR also generate many cell outliers in the microscopy image and nnU-Net failed to successfully recognize the large needle driver in the endoscopy image. In contrast, U-Mamba shows clear advantages across these scenarios, indicating better abilities to capture global context.

\section{Discussion and conclusion}
In this paper, we introduce U-Mamba to address the challenges in modeling long-range dependencies due to the inherent locality of CNNs and the computational complexity of Transformers.
Comprehensive experimental results reveal that U-Mamaba outperforms existing CNN- and Transformer-based segmentation networks across diverse modalities and segmentation targets. 
In particular, U-Mamba has demonstrated superior capabilities in handling objects with heterogeneous appearances, resulting in fewer segmentation outliers. The performance gain is largely attributed to U-Mamba's architecture design, which can simultaneously extract multi-scale local features and capture long-range dependencies.

% why do Transformer-based networks perform so bad
Transformers have demonstrated stronger performance on many natural image segmentation tasks~\cite{Swin}. However, they have not transformed medical image segmentation. This is evident as winning solutions in medical image segmentation competitions still predominantly rely on CNNs\footnote{\url{https://github.com/JunMa11/SOTA-MedSeg}}, such as nnU-Net~\cite{nnUNet} and SegResNet~\cite{SegResNet}.
Interestingly, we also observed that Transformer-based networks underperform relative to CNN-based networks in all our tested scenarios, despite utilizing their original network implementations and training them with the recommended optimizers. 
This could be the requirement of training these networks with multiple GPUs rather than one GPU. However, this would lead to unfair comparisons because all the other networks were trained with only one GPU. Another reason could be that Transformers should be used in a large-scale pre-training and fine-tuning paradigm.  
To maintain consistency and fairness in our evaluation, we trained all networks from scratch and employed identical preprocessing and data augmentation techniques across all experiments. We believe in the value of community collaboration and transparency, and therefore, we have made our implementations of the compared methods publicly available for further examination and use by the research community.

In addition, nnU-Net shows very competitive results on most tasks, but U-Mamba achieves overall better scores. 
We noticed that nnU-Net and U-Mamba have complementary segmentation performance for certain organs in CT and MRI scans, such as the aorta, kidneys, and inferior vena cava as shown in Appendix Table~\ref{tab:organ-3dCT}-\ref{tab:organ-2dMRI}. This suggests the potential for further integrating them via the model ensemble, a strategy often employed to enhance performance in medical image segmentation competitions~\cite{miccai20seg-sota}.

% future improvements: 
While the primary focus of this work is the new segmentation architecture, there are numerous avenues for further enhancement and expansion of U-Mamba. One immediate direction is to leverage large-scale datasets for train U-Mamba, aiming to create readily deployable segmentation tools or provide pre-trained model weights for fine-tuning on data-limited tasks, as exemplified by similar initiatives like TotalSegmentator~\cite{totalsegmentator} and STU-Net~\cite{STUNet}. Furthermore, U-Mamba's design inherently supports integration with advanced techniques, such as strong data augmentations for small datasets, loss functions for highly imbalanced targets, and region-based training for nested objects~\cite{region-training}. In addition, U-Mamba block presents a new opportunity for application in classification and detection networks for better long-range dependency modeling. 
We leave these directions as near future work.

In conclusion, this paper presents a new architecture, U-Mamba, for general-purpose biomedical image segmentation, which integrates the advantages of local pattern recognition from CNNs and global context understanding from Mamba. 
U-Mamba can configure itself in an automatic way for different datasets, making it a versatile and flexible tool for diverse segmentation tasks in biomedical imaging. 
The results suggest that U-Mamba is a promising candidate to serve as the backbone of next-generation biomedical image segmentation networks.

\subsubsection{Acknowledgements} We acknowledge all the authors of the employed public datasets~\cite{simpson2019MSD,TCIA,FLARE22,AMOS22,Endo17,NeurIPS22CellSeg}, allowing the community to use these valuable resources for research purposes. 
We also thank the authors of nnU-Net~\cite{nnUNet} and Mamba~\cite{mamba} for making their valuable code publicly available.

% ---- Bibliography ----
%
% BibTeX users should specify bibliography style 'splncs04'.
% References will then be sorted and formatted in the correct style.
%
\bibliographystyle{splncs04}
\bibliography{ref}

% \newpage

\begin{appendix}

\section*{Appendix}

\begin{table}[htbp]
\caption{Organ-wise segmentation results of 3D models in abdomen CT dataset. The best and the second-best scores for each organ and each metric are highlighted in red and blue, respectively. 
(IVC: Inferior vena cava, RAG: Right adrenal gland, LAG: Left adrenal gland)}\label{tab:organ-3dCT}
\centering
\begin{adjustbox}{width=0.99\textwidth}
\begin{tabular}{lcc|cc|cc|cc|cc|cc}
\hline
\multirow{2}{*}{Organs} & \multicolumn{2}{c|}{nnU-Net}                              & \multicolumn{2}{c|}{SegResNet}                           & \multicolumn{2}{c|}{UNETR}                               & \multicolumn{2}{c|}{SwinUNETR}                           & \multicolumn{2}{c|}{U-Mamba Bot}                                           & \multicolumn{2}{c}{U-Mamba Enc}                                           \\ \cline{2-13} 
                        & DSC                        & NSD                         & DSC                        & NSD                         & DSC                        & NSD                         & DSC                        & NSD                         & DSC                                 & NSD                                  & DSC                                 & NSD                                 \\ \hline
Liver                   & 0.9706                     & \textcolor{blue}{0.9582}    & 0.9521                     & 0.9157                      & 0.9047                     & 0.8125                      & 0.9302                     & 0.8297                      & \textcolor{blue}{0.9713}            & \textcolor{red}{0.9592}              & \textcolor{red}{0.9714}             & 0.9578                              \\
Right kidney            & \textcolor{blue}{0.8738}   & 0.8621                      & 0.8295                     & 0.8027                      & 0.7105                     & 0.6609                      & 0.7540                     & 0.7053                      & 0.8653                              & \textcolor{blue}{0.8640}             & \textcolor{red}{0.8864}             & \textcolor{red}{0.8733}             \\
Spleen                  & 0.9162                     & 0.9038                      & 0.8646                     & 0.8421                      & 0.7869                     & 0.7347                      & 0.8402                     & 0.802                       & \textcolor{red}{0.9357}             & \textcolor{red}{0.9361}              & \textcolor{blue}{0.9241}            & \textcolor{blue}{0.9241}            \\
Pancreas                & 0.8362                     & 0.9139                      & 0.7440                     & 0.8353                      & 0.6010                     & 0.6859                      & 0.6973                     & 0.7191                      & \textcolor{red}{0.8651}             & \textcolor{red}{0.9402}              & \textcolor{blue}{0.8407}            & \textcolor{blue}{0.9157}            \\
Aorta                   & \textcolor{red}{0.9618}    & \textcolor{red}{0.9745}     & 0.9491                     & 0.9558                      & 0.8916                     & 0.8701                      & 0.9383                     & 0.9340                      & \textcolor{blue}{0.9570}            & \textcolor{blue}{0.9727}             & 0.9445                              & 0.9567                              \\
IVC                     & \textcolor{red}{0.8839}    & \textcolor{blue}{0.8776}    & 0.8374                     & 0.8242                      & 0.7592                     & 0.7199                      & 0.8250                     & 0.7834                      & \textcolor{blue}{0.8822}            & \textcolor{red}{0.8784}              & 0.8748                              & 0.8697                              \\
RAG                     & \textcolor{red}{0.8231}    & \textcolor{red}{0.9318}     & 0.7277                     & 0.8579                      & 0.6373                     & 0.7615                      & 0.7466                     & 0.8703                      & \textcolor{blue}{0.8140}            & \textcolor{blue}{0.9236}             & 0.7827                              & 0.8886                              \\
LAG                     & 0.7921                     & 0.8884                      & 0.6851                     & 0.7945                      & 0.4750                     & 0.5866                      & 0.6767                     & 0.7888                      & \textcolor{red}{0.8323}             & \textcolor{red}{0.9258}              & \textcolor{blue}{0.8206}            & \textcolor{blue}{0.9159}            \\
Gallbladder             & 0.7293                     & 0.7331                      & 0.6678                     & 0.6411                      & 0.5321                     & 0.4747                      & 0.5725                     & 0.5320                      & \textcolor{blue}{0.7436}            & \textcolor{blue}{0.7467}             & \textcolor{red}{0.7767}             & \textcolor{red}{0.7786}             \\
Esophagus               & \textcolor{blue}{0.8606}   & \textcolor{blue}{0.9308}    & 0.7927                     & 0.8706                      & 0.6929                     & 0.7826                      & 0.7853                     & 0.8663                      & 0.8523                              & 0.9149                               & \textcolor{red}{0.8689}             & \textcolor{red}{0.9323}             \\
Stomach                 & \textcolor{blue}{0.8902}   & 0.9033                      & 0.8207                     & 0.8302                      & 0.7128                     & 0.6903                      & 0.7656                     & 0.6981                      & \textcolor{red}{0.8931}             & \textcolor{red}{0.9076}              & 0.8856                              & \textcolor{blue}{0.9034}            \\
Duodenum                & \textcolor{blue}{0.7561}   & \textcolor{blue}{0.8830}    & 0.6388                     & 0.7961                      & 0.4951                     & 0.6932                      & 0.6042                     & 0.7531                      & \textcolor{red}{0.7794}             & \textcolor{red}{0.8930}              & 0.7534                              & 0.8692                              \\
Left kidney             & \textcolor{red}{0.9050}    & \textcolor{red}{0.9035}     & 0.7949                     & 0.7686                      & 0.6724                     & 0.6324                      & 0.7367                     & 0.6799                      & 0.8963                              & \textcolor{blue}{0.9019}             & \textcolor{blue}{0.8991}            & 0.8889                              \\ \hline
Average                 & 0.8615 & 0.8972 & 0.7926 & 0.8258 & 0.6824 & 0.7004 & 0.7594 & 0.7663 & \textcolor{red}{0.8683} & \textcolor{red}{0.9049} & \textcolor{blue}{0.8638} & \textcolor{blue}{0.8980} \\ \hline
\end{tabular}
\end{adjustbox}
\end{table}

\begin{table}[htbp]
\caption{Organ-wise segmentation results of 3D models in abdomen MRI dataset. The best and the second-best scores for each organ and each metric are highlighted in red and blue, respectively. 
(IVC: Inferior vena cava, RAG: Right adrenal gland, LAG: Left adrenal gland)}\label{tab:organ-3dMRI}
\centering
\begin{adjustbox}{width=0.99\textwidth}
\begin{tabular}{lll|ll|ll|ll|ll|ll}
\hline
\multirow{2}{*}{Organs} & \multicolumn{2}{c|}{nnU-Net}                        & \multicolumn{2}{c|}{SegResNet}                     & \multicolumn{2}{c|}{UNETR}                         & \multicolumn{2}{c|}{SwinUNETR}                     & \multicolumn{2}{c|}{U-Mamba Bot}                   & \multicolumn{2}{c}{U-Mamba Enc}                   \\ \cline{2-13} 
                        & \multicolumn{1}{c}{DSC} & \multicolumn{1}{c|}{NSD} & \multicolumn{1}{c}{DSC} & \multicolumn{1}{c|}{NSD} & \multicolumn{1}{c}{DSC} & \multicolumn{1}{c|}{NSD} & \multicolumn{1}{c}{DSC} & \multicolumn{1}{c|}{NSD} & \multicolumn{1}{c}{DSC} & \multicolumn{1}{c|}{NSD} & \multicolumn{1}{c}{DSC} & \multicolumn{1}{c}{NSD} \\ \hline
Liver                   & \textcolor{red}{0.9735} & \textcolor{blue}{0.9747} & 0.9654                  & 0.9590                   & 0.9344                  & 0.8905                   & 0.9627                  & 0.9479                   & \textcolor{blue}{0.9732}& \textcolor{red}{0.9753}  & 0.9717                  & 0.9704                  \\
Right kidney            & \textcolor{red}{0.9625} & 0.9747                   & 0.9390                  & 0.9588                   & 0.8266                  & 0.8228                   & 0.9355                  & 0.9390                   & 0.9594                  & \textcolor{blue}{0.9766} & \textcolor{blue}{0.9621}& \textcolor{red}{0.9800} \\
Spleen                  & 0.9131                  & 0.9201                   & 0.9023                  & 0.9069                   & 0.8617                  & 0.8280                   & 0.9134                  & 0.9019                   & \textcolor{blue}{0.9383}& \textcolor{blue}{0.9417} & \textcolor{red}{0.9442} & \textcolor{red}{0.9468} \\
Pancreas                & \textcolor{blue}{0.8639}& 0.9553                   & 0.8249                  & 0.9228                   & 0.7271                  & 0.8268                   & 0.7750                  & 0.8742                   & \textcolor{red}{0.8650} & \textcolor{red}{0.9565}  & 0.8623                  & \textcolor{blue}{0.9560}\\
Aorta                   & 0.9327                  & \textcolor{blue}{0.9572} & \textcolor{blue}{0.9331}& 0.9556                   & 0.8524                  & 0.8652                   & 0.8907                  & 0.9055                   & 0.9243                  & 0.9479                   & \textcolor{red}{0.9467} & \textcolor{red}{0.9711}                 \\
IVC                     & 0.8192                  & 0.8698                   & \textcolor{blue}{0.8279}& \textcolor{blue}{0.8762} & 0.7286                  & 0.7605                   & 0.7844                  & 0.8271                   & \textcolor{red}{0.8322} & \textcolor{red}{0.8812}  & 0.8230                  & 0.8740                  \\
RAG                     & 0.6238                  & \textcolor{blue}{0.8047}                   & 0.6066                  & 0.7778                   & 0.4539                  & 0.6257                   & 0.5540                  & 0.7316                   & \textcolor{blue}{0.6287}                  & 0.8010                   & \textcolor{red}{0.6438}                  & \textcolor{red}{0.8197}                  \\
LAG                     & 0.7021                  & 0.8474                   & 0.6779                  & 0.8301                   & 0.4735                  & 0.6143                   & 0.4786                  & 0.6255                   & \textcolor{blue}{0.7225}                  & \textcolor{blue}{0.8588}                   & \textcolor{red}{0.7419}                  & \textcolor{red}{0.8713}                  \\
Gallbladder             & 0.7720                  & 0.7551                   & 0.7872                  & 0.7581                   & 0.4833                  & 0.4095                   & 0.6537                  & 0.6185                   & \textcolor{blue}{0.8284}                  & \textcolor{blue}{0.8279}                   & \textcolor{red}{0.8503}                  & \textcolor{red}{0.8449}                  \\
Esophagus               & 0.7489                  & 0.9071                   & 0.7408                  & 0.8976                   & 0.5655                  & 0.7395                   & 0.6477                  & 0.8038                   & \textcolor{red}{0.7955}                  & \textcolor{red}{0.9387}                   & \textcolor{blue}{0.7890}                  & \textcolor{blue}{0.9369}                  \\
Stomach                 & \textcolor{blue}{0.8260}                  & 0.8595                   & 0.7609                  & 0.7989                   & 0.6697                  & 0.6919                   & 0.7198                  & 0.7553                   & 0.8257                  & \textcolor{blue}{0.8597}                   & \textcolor{red}{0.8283}                  & \textcolor{red}{0.8679}                  \\
Duodenum                & 0.7008                  & 0.8892                   & 0.6634                  & 0.8720                   & 0.4878                  & 0.7495                   & 0.5921                  & 0.8148                   & \textcolor{red}{0.7309}                  & \textcolor{red}{0.9078}                   & \textcolor{blue}{0.7232}                  & \textcolor{blue}{0.8991}                  \\
Left kidney             & 0.9637                  & 0.9805                   & 0.9609                  & 0.9798                   & 0.8622                  & 0.8479                   & 0.9266                  & 0.9377                   & \textcolor{red}{0.9649}                  & \textcolor{red}{0.9842}                   & \textcolor{blue}{0.9647}                  & \textcolor{red}{0.9842}                  \\ \hline
Average                 & 0.8309                  & 0.8996                   & 0.8146                  & 0.8841                   & 0.6867                  & 0.7440                   & 0.7565                  & 0.8218                   & \textcolor{blue}{0.8453}         & \textcolor{blue}{0.9121}         & \textcolor{red}{0.8501}         & \textcolor{red}{0.9171}         \\ \hline
\end{tabular}
\end{adjustbox}
\end{table}

\begin{table}[htbp]
\caption{Organ-wise segmentation results of 2D models in abdomen MRI dataset. The best and the second-best scores for each organ and each metric are highlighted in red and blue, respectively. 
(IVC: Inferior vena cava, RAG: Right adrenal gland, LAG: Left adrenal gland)}\label{tab:organ-2dMRI}
\centering
\begin{adjustbox}{width=0.99\textwidth}
\begin{tabular}{lll|ll|ll|ll|ll|ll}
\hline
\multirow{2}{*}{Organs} & \multicolumn{2}{c|}{nnU-Net}                        & \multicolumn{2}{c|}{SegResNet}                     & \multicolumn{2}{c|}{UNETR}                         & \multicolumn{2}{c|}{SwinUNETR}                     & \multicolumn{2}{c|}{U-Mamba Bot}                   & \multicolumn{2}{c}{U-Mamba Enc}                   \\ \cline{2-13} 
                        & \multicolumn{1}{c}{DSC} & \multicolumn{1}{c|}{NSD} & \multicolumn{1}{c}{DSC} & \multicolumn{1}{c|}{NSD} & \multicolumn{1}{c}{DSC} & \multicolumn{1}{c|}{NSD} & \multicolumn{1}{c}{DSC} & \multicolumn{1}{c|}{NSD} & \multicolumn{1}{c}{DSC} & \multicolumn{1}{c|}{NSD} & \multicolumn{1}{c}{DSC} & \multicolumn{1}{c}{NSD} \\ \hline
Liver                   & 0.9478                  & 0.9286                   & 0.9321                  & 0.9051                   & 0.8988                  & 0.8173                   & 0.9484                  & 0.9108                   & \textcolor{blue}{0.9534}                  & \textcolor{blue}{0.9319}                   & \textcolor{red}{0.9579}                  & \textcolor{red}{0.9390}                  \\
Right kidney            & 0.8782                  & 0.8749                   & 0.8957                  & 0.8802                   & 0.6974                  & 0.6606                   & 0.8598                  & 0.8391                   & \textcolor{blue}{0.9039}                  & \textcolor{blue}{0.8922}                   & \textcolor{red}{0.9070}                  & \textcolor{red}{0.9043}                  \\
Spleen                  & 0.9043                  & 0.8956                   & 0.8793                  & 0.8573                   & 0.7182                  & 0.6612                   & 0.8922                  & 0.8649                   & \textcolor{blue}{0.9065}                  & \textcolor{blue}{0.8992}                   & \textcolor{red}{0.9137}                  & \textcolor{red}{0.9065}                  \\
Pancreas                & 0.7148                  & 0.8398                   & 0.6998                  & 0.8257                   & 0.5679                  & 0.6990                   & 0.7049                  & 0.8260                   & \textcolor{blue}{0.7376}                  & \textcolor{blue}{0.8577}                   & \textcolor{red}{0.7500}                  & \textcolor{red}{0.8715}                  \\
Aorta                   & 0.9214                  & 0.9465                   & \textcolor{red}{0.9333}                  & \textcolor{red}{0.9582}                   & 0.8294                  & 0.8401                   & 0.8750                  & 0.8929                   & \textcolor{blue}{0.9257}                  & \textcolor{blue}{0.9571}                   & 0.9122                  & 0.9409                  \\
IVC                     & 0.7567                  & 0.8045                   & 0.7527                  & 0.8003                   & 0.6310                  & 0.6703                   & 0.7105                  & 0.7536                   & \textcolor{red}{0.7653}                  & \textcolor{blue}{0.8136}                   & \textcolor{blue}{0.7631}                  & \textcolor{red}{0.8147}                  \\
RAG                     & 0.4960                  & 0.6743                   & 0.4746                  & 0.6755                   & 0.3478                  & 0.5243                   & 0.4552                  & 0.6162                   & \textcolor{red}{0.5513}                  & \textcolor{red}{0.7338}                   & \textcolor{blue}{0.5371}                  & \textcolor{blue}{0.7293}                  \\
LAG                     & 0.5419                  & 0.7023                   & 0.5542                  & 0.7089                   & 0.3471                  & 0.4764                   & 0.4758                  & 0.6221                   & \textcolor{red}{0.5682}                  & \textcolor{red}{0.7283}                   & \textcolor{blue}{0.5642}                  & \textcolor{blue}{0.7163}                  \\
Gallbladder             & \textcolor{red}{0.6807}                  & \textcolor{blue}{0.6371}                   & 0.6256                  & 0.5959                   & 0.3103                  & 0.2557                   & 0.6034                  & 0.5581                   & 0.6664                  & 0.6355                   & \textcolor{blue}{0.6762}                  & \textcolor{red}{0.6413}                  \\
Esophagus               & 0.6964                  & 0.8605                   & 0.6756                  & 0.8542                   & 0.5648                  & 0.7602                   & 0.6347                  & 0.8216                   & \textcolor{red}{0.7042}                  & \textcolor{blue}{0.8633}                   & \textcolor{blue}{0.7025}                  & \textcolor{red}{0.8669}                  \\
Stomach                 & 0.7351                  & 0.7622                   & 0.7134                  & 0.7437                   & 0.5583                  & 0.5756                   & 0.6739                  & 0.6945                   & \textcolor{blue}{0.7420}                  & \textcolor{blue}{0.7636}                   & \textcolor{red}{0.7680}                  & \textcolor{red}{0.7888}                  \\
Duodenum                & 0.5204                  & 0.7842                   & 0.4784                  & 0.7450                   & 0.3054                  & 0.6156                   & 0.4448                  & 0.7335                   & \textcolor{blue}{0.5295}                  & \textcolor{blue}{0.7890}                   & \textcolor{red}{0.5516}                  & \textcolor{blue}{0.7952}                  \\
Left kidney             & 0.8911                  & 0.8881                   & 0.898                   & 0.8943                   & 0.6944                  & 0.6459                   & 0.8580                  & 0.8361                   & \textcolor{red}{0.9100}                  & \textcolor{blue}{0.9058}                   & \textcolor{blue}{0.9084}                  & \textcolor{red}{0.9106}                  \\ \hline
Average                 & 0.7450                  & 0.8153                   & 0.7317                  & 0.8034                   & 0.5747                  & 0.6309                   & 0.7028                  & 0.7669                   & \textcolor{blue}{0.7588}         & \textcolor{blue}{0.8285}          & \textcolor{red}{0.7625}         & \textcolor{red}{0.8327}         \\ \hline
\end{tabular}
\end{adjustbox}
\end{table}

\end{appendix}

\end{document}